# Anomalous Aharonov-Bohm Interference in the Presence of Edge Reconstruction

Sourav Biswas, Hemanta Kumar Kundu, Rajarshi Bhattacharyya, Vladimir Umansky
and Moty Heiblum[*]

*Braun Center for Submicron Research, Department of Condensed Matter Physics, Weizmann Institute of Science, Rehovot 7610001, Israel*

**Interferometry is a vital tool for studying fundamental features in the quantum Hall effect (QHE). For instance, Aharonov-Bohm (AB) interference in a quantum Hall interferometer can probe the wave-particle duality of electrons and quasiparticles. Here, we report an unusual AB interference in a quantum Hall Fabry-Pérot interferometer (FPI), whose Coulomb interactions were suppressed with a grounded drain in the interior bulk of the FPI. In a descending filling factor from $\nu = 3$ to $\nu \approx \frac{5}{3}$, the magnetic field periodicity, which corresponded to a single 'flux quantum,' agreed accurately with the enclosed area of the FPI. However, in the filling range, $\nu \approx \frac{5}{3}$ to $\nu = 1$, the field periodicity increased markedly, apriori suggesting a drastic shrinkage of the AB area. Moreover, the modulation gate voltage periodicity decreased abruptly at this range. We attribute these unexpected observations to a ubiquitous edge reconstruction, leading to *dynamical* area changing with the field and a modified modulation gate-edge capacitance. These results are reproducible and support future interference experiments with a QHE-FPI.**

A two-dimensional electron gas (2DEG), subjected to a strong perpendicular magnetic field, exhibits quantized plateaus of the electrical Hall conductance, $\sigma_{xy} = \nu \frac{e^2}{h}$, where $e$ is the electron charge, $h$ is the Planck constant, and $\nu$ is the filling factor (integer or fraction). In a quantum Hall (QH) system, gapless, chiral, 1D-like edge modes carry the current while the bulk is insulating[1-3]. Carriers can be electrons or fractionally charged quasiparticles[4-6], with the latter obeying anyonic exchange statistics, with the quasiparticles exchange phase $\pi/m$ at a filling $\nu = 1/m$[7-9]. Studying the edge modes allows for discerning the properties of the bulk (due to 'bulk-edge correspondence'). Examples are the studies of shot noise[10], thermal transport[11], and interference[12].

Interferometry with QH edge modes provides an excellent tool for investigating the wave nature of electrons and quasiparticles, their coherence, and statistics. The customarily employed interferometers in the QH regime are the multi-path Fabry-Pérot interferometer (FPI)[13-20] and the two-path Mach-Zehnder interferometer (MZI)[21-23]. While the MZI has a built-in metallic drain, thus operating in the Aharonov-Bohm (AB) regime, an unscreened FPI is affected by Coulomb interactions[24,25], which tend to mask the AB interference. Screening these interactions with parallel conducting layers[19], a metallic top gate[13,14], or via draining charges by an internal grounded drain[15,17] reveal AB interference. Anyonic statistics had been recently observed in a screened FPI[20].

Here, we studied interference in FPIs with an internal grounded drain in the fillings range $\nu = 3 - 1$. As detailed below, an anomalous interference of the outermost integer edge mode appears concomitantly with the emergence of edge reconstruction at $\nu \approx 5/3$.[26-28]

A ubiquitous FPI consists of two partitioning quantum point contacts (QPCs) serving as beam splitters, a gated (or etched) confined bulk, and a 'modulation gate' [Supplementary Fig. S1]. The latter depletes the bulk locally, thus changing the area enclosed by the interfering edge mode, $\delta A = \alpha \delta V_{MG}$, with $\alpha$ a function of the gate-edge capacitance and the carrier density. The AB flux periodicity of the interfering integer edge mode is the flux quantum, $\Phi_0 = h/e$, with a winding phase $\varphi_{AB} = 2\pi BA/\Phi_0$, where $A$ is the enclosed area and $B$ the magnetic field[29]. For weakly backscattering QPCs, the transmission probability is given by, $T_{FPI} = |\tau|^2 = t_l^2 t_r^2 (1 + r_l^2 r_r^2 \cos \varphi_{AB})$, where $t_l^2$ ($t_r^2$) and $r_l^2$ ($r_r^2$) are the transmission and reflection probabilities of the left (right) QPC, respectively. In the fractional regime, the AB phase is modified, $\varphi_{AB} = \frac{e^*}{e} 2\pi BA/\Phi_0$, leading to periodicities: $\Delta B = \frac{e}{e^*}\Phi_0/A$ and $\Delta A = \frac{e}{e^*}\Phi_0/B$.

The FPIs were fabricated in a high-mobility 2DEG embedded in GaAs-AlGaAs heterostructure at a depth $\sim 100$ nm below the surface, with two different electron densities $(1.7 \: \& \: 1.1) \times 10^{11}$ cm$^{-2}$ and an enclosed area between 5 μm$^2$ and 17 μm$^2$ [Fig. 1(a) and Supplementary Fig. S2]. The internal grounded drain dimensions varied between $450 \times 450$ nm$^2$ and $800 \times 800$ nm$^2$, respectively. Differential conductance was measured by applying a small AC voltage of 1 μ$V_{RMS}$ at 900 KHz at 10 mK. The signal was amplified by a cooled preamplifier (at 1.5 K) cascaded by a room-temperature amplifier.

We start with interfering the outermost edge mode at fillings $\nu = 3$ and $\nu = 2$ in an FPI with an area of ~10.25 μm². The conductance plot in the $B - V_{MG}$ plane[29] has the familiar AB-type *pajama* pattern [Figs. 1(b) & 1(c)]; clearly proving a typical AB interference. The AB periodicities at $\nu = 2$ are: $\Delta B = 6.4$ G & $\Delta V_{MG} = 6.58$ mV, corresponding to electron interference with a flux periodicity of $\Phi_0$. However, at $\nu = 3$ we find: $\Delta B = 3.2$ G & $\Delta V_{MG} = 5.55$ mV corresponding to a flux periodicity of $\frac{\Phi_0}{2}$; being the known (but not understood) pairing phenomenon of the interfering electrons, with $e^* \approx 2e$.[15,30] Throughout the filling range $\nu = 3 - 5/3$, the corresponding AB area is $A = \frac{e}{e^*}\frac{\Phi_0}{\Delta B} = 6.4$ μm², suggesting gate-induced lateral depletion of some $\approx 350$ nm.

At the filling factor $\nu \approx 5/3$, a dramatic change in the interference pattern takes place. Monitoring the interference in a progressively declining filling factor between $\nu \approx 5/3$ and $\nu = 1$, the field periodicity increases from $\Delta B = 6.4$ G to $\Delta B = 18.6$ G, respectively, see Figs. 2(a) & 2(b) and Supplementary Fig. S4 for the AB pajama at $\nu = 1$. Also, the modulation gate voltage periodicity decreases unexpectedly with lower filling in the same filling range.

The dramatic change in the field periodicity in the range $\nu = 5/3 - 1$ suggests apriori a monotonic shrinkage of the area $\left(\frac{\Phi_0}{\Delta B}\right)$ – by a factor of around three at $\nu = 1$ [Fig. 3(a)]. However, since FPIs with different lithographic areas exhibit similar behavior, i.e., an approximately similar factor of increase in $\Delta B$, see Fig. 3(a), such apparent 'area change' cannot occur. What is so unique in $\nu \approx 5/3$ and lower filling that leads to such abnormal behavior?

We recently found an emergent edge reconstruction of the outermost edge mode starting at $\nu \approx 5/3$ and persisting until $\nu = 1$. The integer filling $\nu = 1$ was reconstructed to two fractional states; an outermost $\nu = 1/3$ and an inner $\nu = 2/3$[26,31,32], with $\frac{e^2}{3h}$ conductance plateau observed in the transmission of a QPC [Fig. 2(c)]. Beyond an equilibration distance, an upstream bosonic neutral mode is born [Fig. 2(d)]. Approaching $\nu = 1$, edge reconstruction and the excitation of the neutral mode strengthen. Based on these observations, we propose a mechanism explaining anomalous behavior and the apparent area shrinkage.

The enclosed flux evolution is $\Delta \Phi = \Delta B * A + B * \Delta A$, with $A$ as the enclosed area. Assume that with increasing $B$ the enclosed area does not remain constant, leading to $\frac{1}{A}\frac{\Delta \Phi}{\Delta B} = 1 + \frac{B}{\Delta B}\frac{\Delta A}{A}$. For filling $\nu > 5/3$, where $\Delta B$ is constant, namely, $A * \Delta B = \Phi_0$ and $\frac{1}{A}\frac{\Delta \Phi}{\Delta B} = 1$, with $\Delta A = 0$ – as expected. However, for $\nu < 5/3$, since the periodicity of $\Delta B$ increases monotonically with $B$, a *dynamic* reduction in the area, $\Delta A$, must occur in each period of one flux quantum. The estimated magnitude of $\Delta A$, with the measured values of $B$ and $\Delta B$, is shown in Fig. 3(c). The relative change in the area is tiny, $\frac{\Delta A}{A} \sim 10^{-4}$, resulting from an inward shift δ of the circulating edge mode, $\delta = \frac{\Delta A}{4\sqrt{A}}$, which is of the order of ~0.1 nm [Fig. 3(c)]. It is remarkable how a minute (dynamic) change in $\delta$ strongly affects the AB interference periodicity. See Supplementary Section-I for further detailed analysis.

At filling $\nu \approx 5/3$, due to edge reconstruction an incompressible strip is formed around the outermost edge, with fillings $\nu = 1/3$ and $\nu = 2/3$ [23]. With increasing the field (with further lower filling), this strip continues to widen and thus in field scanning, the dynamic area ($\Delta A$) plays a significant role. Approaching $\nu = 1$, the increasing reconstruction strength and wider strip lead to the proportional increase in $\delta$ with $B$. The higher (lower) $\delta$ for a smaller (larger) size device is consistent with the larger (smaller) $\Delta B$, i.e., one needs to scan a longer (shorter) range of $B$ for one flux quantum resulting in the wider (narrower) reconstructed strip.

Similarly, the dependence of $\frac{1}{\Delta V_{MG}}$ on $B$, with $\Delta V_{MG}$ the voltage required to change the AB flux by one flux quantum changes abruptly at $\nu \approx 5/3$ (with $\Delta \Phi = B * \Delta A$) [Fig. 3(b)]. The incremental depleted charge, $\Delta Q = n \Delta A = C_{MG} \Delta V_{MG}$, or equivalently $\frac{1}{\Delta V_{MG}} = \frac{C_{MG}}{n} * \frac{B}{\Phi_0}$, where $n$ is the charge density, $C_{MG}$ is the gate-edge capacitance, and $\alpha = \frac{C_{MG}}{n}$. Using the measured $\Delta V_{MG}$, $\alpha$ is plotted as a function of $B$ in Fig. 3(d). While $\alpha$ is nearly constant (as expected) for $\nu > 5/3$, it increases at fillings $\nu < 5/3$. The larger $\alpha$ is a result of the lower charge density in the reconstructed lower filling (e.g., $\nu = 1/3$) region. The latter getting stronger leads to the continuous increase in $\alpha$ with $B$.

Since edge reconstruction can depend on the sharpness of the edge potential, we repeated these measurements in an FPI, which allows tunable pinching [Fig. 4(a)]. We tested the frequency $\frac{\Phi_0}{\Delta B}$ with a more substantial confining potential (via $V_G$) in the range $\nu = 3$ to $\nu = 1$ [Fig. 4(b)]. We find the following: ***i.*** The AB area decreases drastically with increasing gate depletion; ***ii.*** The anomaly in the periodicity for fillings $\nu < 5/3$ persists and suggests that edge reconstruction and neutral modes cannot be fully quenched. In the Supplementary Section, Fig. S7, we describe similar experiments performed in an MZI, where there is nearly no change in the AB area. Indeed, the two interfering trajectories in an MZI shift in the same direction with $B$, while in the FPI the trajectories move in the opposite direction.

The Fabry-Pérot interferometer is the simplest and thus frequently employed in the quantum Hall regime. Its primary disadvantages are the unavoidable charging energy, which affects the 'Aharonov-Bohm interference area' with changing magnetic flux, and the ill-defined area due to the geometry[33]. When a ubiquitous edge reconstruction occurs, topological or trivial, electron interference evolves unpredictably. In contrast to the expected periodicities in the magnetic field and gate voltage, a gradual increase in the periodicity is observed with an increasing magnetic field, accompanied by an abnormal decrease in the gate voltage periodicity.

We thank Mitali Banerjee, Dima E. Feldman, Yuval Gefen, Bertrand I. Halperin, Bernd Rosenow, and Ady Stern for useful discussions. S.B. acknowledges Abhay Kumar Nayak for his technical help. We acknowledge the continuous support of the Sub-Micron Center staff. M.H. acknowledges the support of the European Research Council under the European Union's Horizon 2020 research and innovation program (grant agreement number 833078) and the partial support of the Minerva Foundation under grant number 713534.

S.B., R.B., and H.K.K. fabricated the devices and performed the measurements. S.B., H.K.K., and M.H. analyzed the data, discussed understanding the results, and wrote the paper. V.U. grew the GaAs heterostructures. M.H. supervised the experiments.

[*]Corresponding author: moty.heiblum@weizmann.ac.il

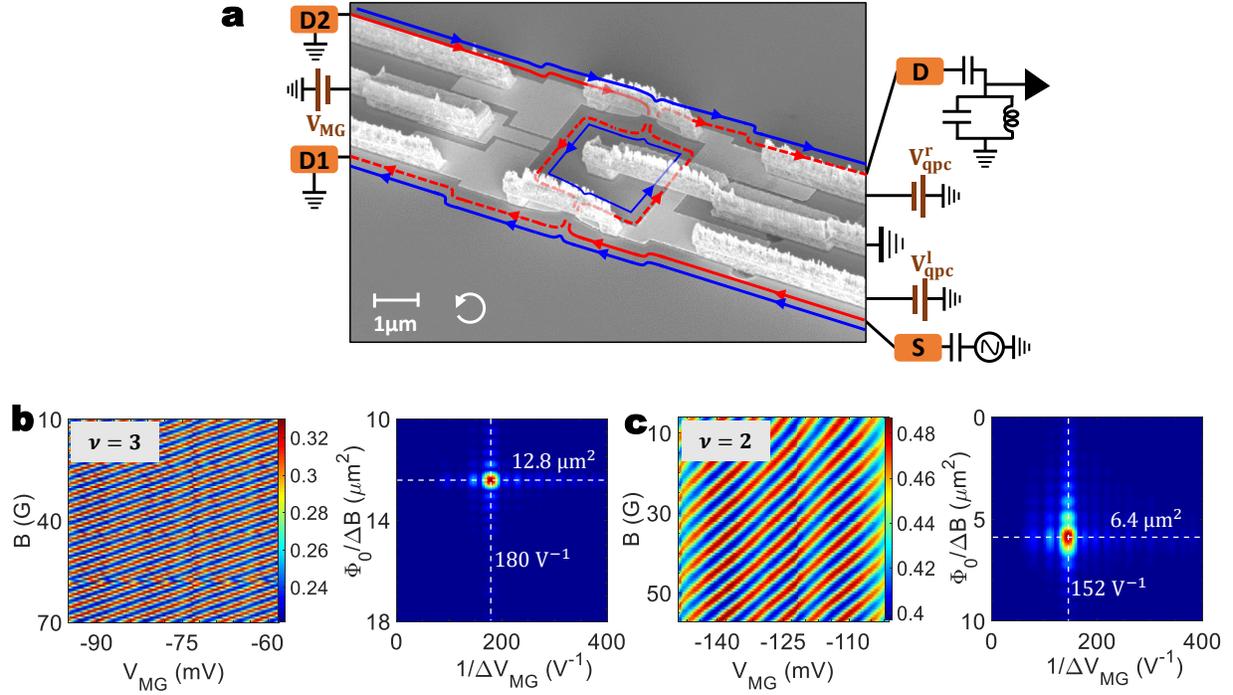

**Figure 1 | Device structure and Aharonov-Bohm oscillations at $\nu = 3, 2$.** (a) Scanning electron micrograph image of a Fabry-Pérot interferometer with a grounded drain at the center bulk. Standard ohmic contacts of alloyed Ni-Au-Ge served as source (S) and drain (D), and thin metallic PdAu-Au served as gates. The split gates as QPCs were 450 nm apart and the modulation gate was 300 nm wider. The lithographic internal area of the device is $10.25\ \mu m^2$. (b, c) Typical Aharonov-Bohm pajama in $B - V_{MG}$ plane at filling factor $\nu = 3$ ($B = 2.33$ T) and $\nu = 2$ ($B = 3.6$ T), when the outer edge is weakly partitioned (the transmission probability $t_{l,r}^2 \approx 0.93$) by the QPCs. Respective first Fourier transformation with the unique frequencies in $B$ and $V_{MG}$ are shown on the right. The values of $\Delta B$ and $\Delta V_{MG}$ follow the usual AB interference equation resulting in the flux periodicity of $\Phi_0$ at $\nu = 2$, and $\frac{\Phi_0}{2}$ at $\nu = 3$.

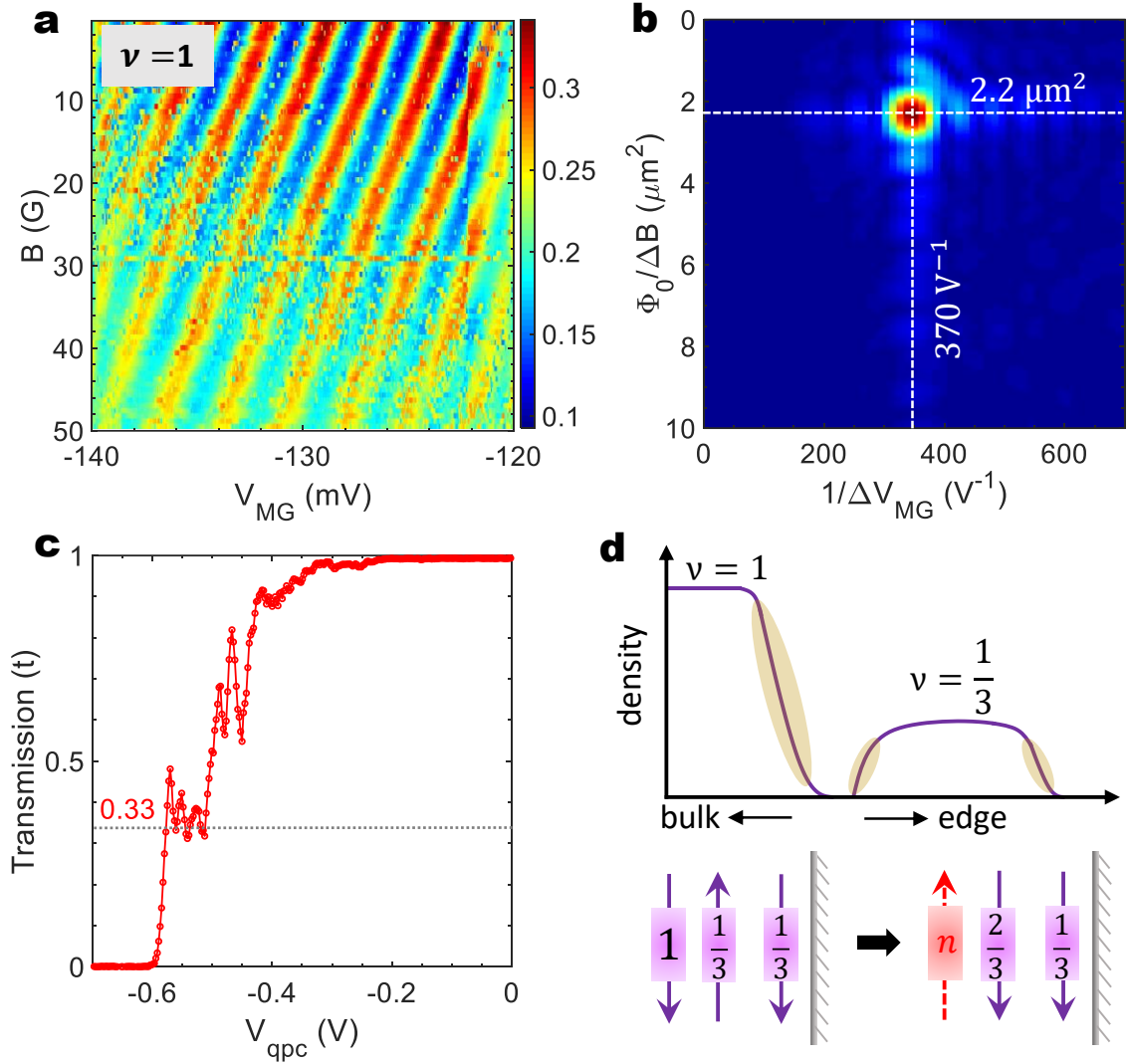

**Figure 2 | Aharonov-Bohm interference and the reconstructed edge at $\nu = 1$.** (a) The characteristic pajama with the Aharonov-Bohm (AB) slope at filling factor $\nu = 1$ ($B = 6.4$ T). The QPCs' transmissions are $t_{l,r}^2 \approx 0.46$. (b) The first Fourier transformation shows a single peak with periodicity, $\Delta V_{MG} = 2.7$ mV & $\Delta B = 18.6$ G, drastically different than what is expected from $\nu = 2$. Comparing with the periodicities at $\nu = 2$ and using the conventional equation, the obtained AB flux periodicity at $\nu = 1$ is $2.9\Phi_0$ (in $B$-dependence) but $0.73\Phi_0$ (in $V_{MG}$-dependence) – hence unacceptable. (c) The conductance as the QPC transmission shows a plateau of $1/3$ due to edge reconstruction. Interference is measured at different transmission values with the QPCs weakly and strongly pinched (see Supplementary Fig. S4). (d) Schematic representation of the reconstruction of the outer edge mode. The creation of $1/3$ density near the edge leads to two counterpropagating $1/3$ modes. The modes' equilibration leads to two downstream $2/3$ and $1/3$

modes, and an upstream neutral mode such that the net thermal Hall conductance $\kappa_{xy}$ remains conserved.

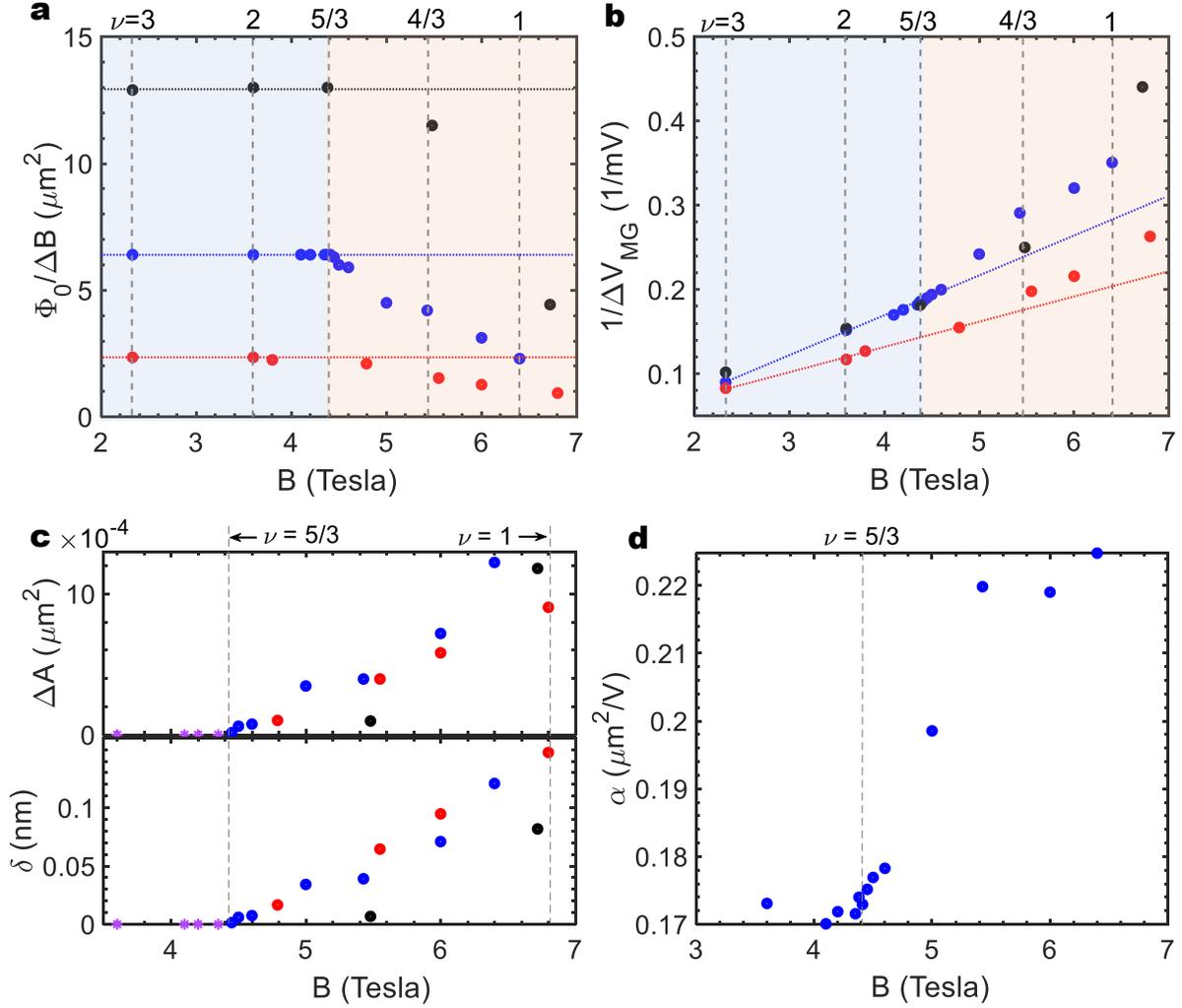

**Figure 3 | Anomalous periodicity of interference at $\nu < 5/3$. (a,b)** Variation of field- and modulation gate-periodicities with filling factors ranging from $\nu = 3$ to $1$ for three different devices of lithographic internal area $17.2~\mu m^2$ (black), $10.25~\mu m^2$ (blue), and $4.87~\mu m^2$ (red). The estimated apparent area $\left(\frac{\Phi_0}{\Delta B}\right)$ as a function of the magnetic field ($B$) shows a monotonic decrease below the filling $\nu \approx 5/3$. Similarly, the frequency $\left(\frac{1}{\Delta V_{MG}}\right)$ with $B$ also diverts from the usual linear slope of electrons' interference below $\nu \approx 5/3$. **(c)** The calculated dynamic area $\Delta A$ from the obtained $\Delta B$ and the amount of inward shift $\delta$ of the edge mode. For $\nu > 5/3$, $\Delta A = 0$ and $\delta = 0$ (purple star), while they increase with $B$ in the range $\nu = 5/3$ to $1$ (solid circle). **(d)** The mutual capacitance $\alpha$, obtained from $\Delta V_{MG}$ values, for a device represented in blue dots showing its' unusual variation below $\nu \approx 5/3$.

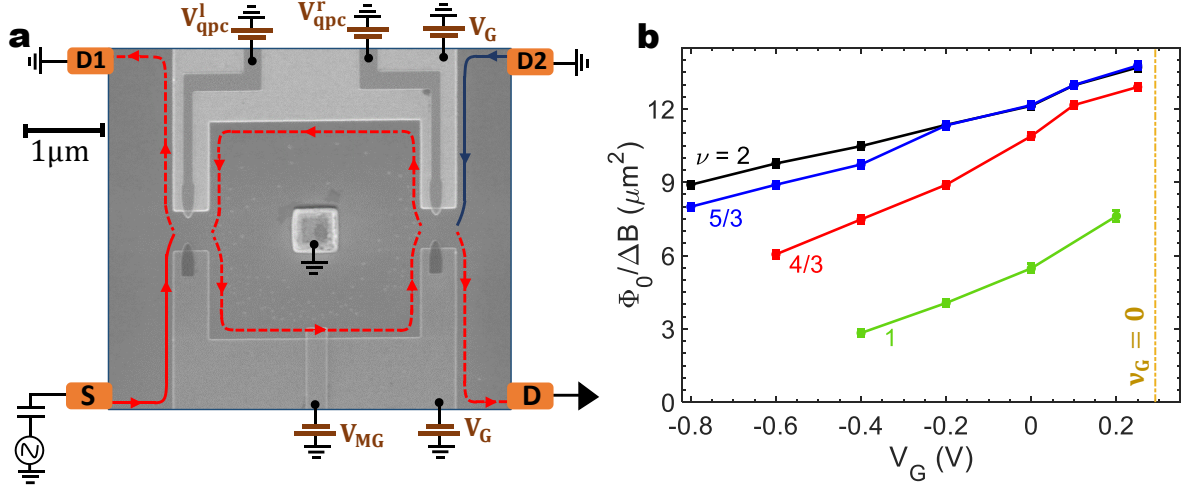

**Figure 4 | Interferometer with variable edge potential and the magnetic field periodicity. (a)** Scanning electron micrograph image of a fully gated Aharonov-Bohm Fabry-Pérot interferometer (FPI) with the internal lithographic area of $14.2\ \mu m^2$, fabricated in the lower density 2DEG. The split gates (QPCs) are separated from the rest of the confining gates. Hafnium Oxide separates the ohmic contacts (shown in yellow) from the gates and between the successive gate layers. The gate voltage $V_G$ tunes the depletion and thus the edge potential. The QPC gate voltages $V_{qpc}$ independently induce backscattering to the edge mode. **(b)** Obtained apparent area $\left(\frac{\Phi_0}{\Delta B}\right)$ from the magnetic field-dependent conductance oscillations at different depletion strengths of the FPI at different filling factors. More details of interference traces and pajamas are presented in Supplementary Fig. S5 and Fig. S6. $V_G = 0.27$ V is the value at which the filling under the gate just becomes zero, $\nu_G = 0$ (shown by the dashed yellow line). With lower $V_G$, the effective area reduction is significant (e.g., ~36% reduces for $V_G = -0.8$ V at $\nu = 2$). Along any vertical cut (constant $V_G$ line), the anomaly in the value of $\frac{\Phi_0}{\Delta B}$ below the filling of $\approx 5/3$ is clearly realized.

# Supplementary Data

**"Anomalous Aharonov-Bohm Interference in the Presence of Edge Reconstruction"**

Sourav Biswas, Hemanta Kumar Kundu, Rajarshi Bhattacharyya, Vladimir Umansky
and Moty Heiblum[*]

*Braun Center for Submicron Research, Department of Condensed Matter Physics, Weizmann Institute of Science, Rehovot 7610001, Israel*

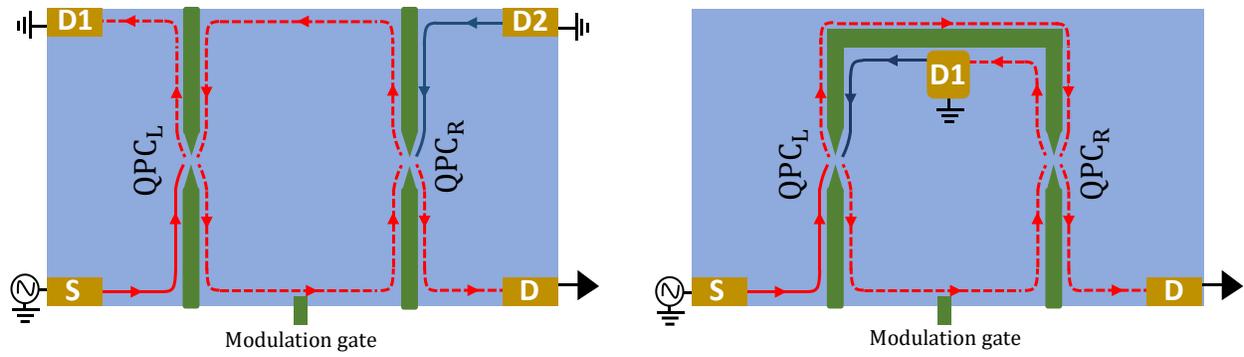

**Figure S1 | Fabry-Pérot and Mach-Zehnder interferometer.** Schematics of a multi-path Fabry-Pérot (left) and a two-path Mach-Zehnder interferometer (right). The two QPCs act as the beam splitters of the electron waves. The modulation gate at the periphery tunes the area of the interferometer.

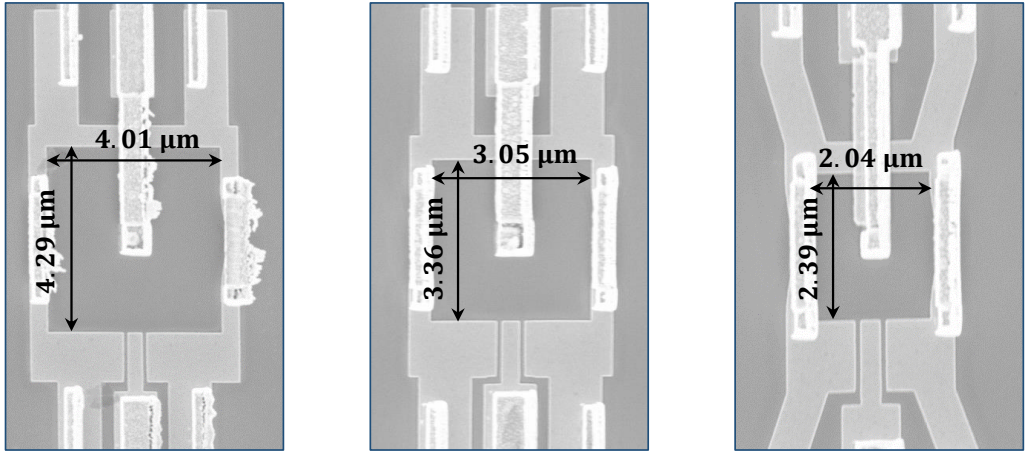

**Figure S2 | Images of our Fabry-Pérot interferometers.** Scanning electron micrograph images of different-sized interferometers investigated. Calculated lithographic internal areas (from left to right) are 17.2 µm², 10.25 µm², and 4.87 µm².

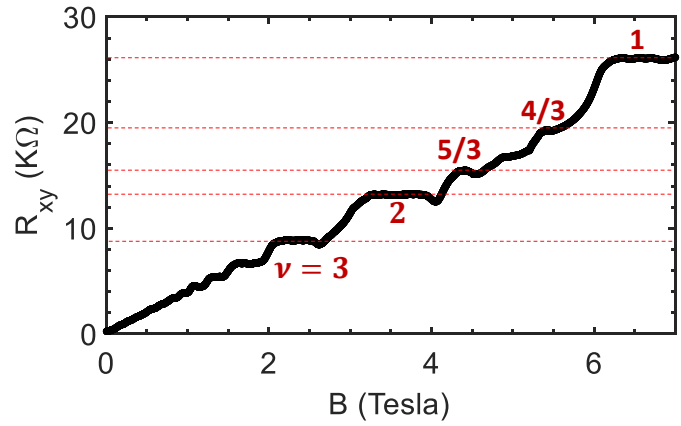

**Figure S3 | Hall measurement.** Hall resistance $R_{xy}$ as a function of the magnetic field $B$. The 2DEG electron density is $1.7 \times 10^{11}$ cm$^{-2}$.

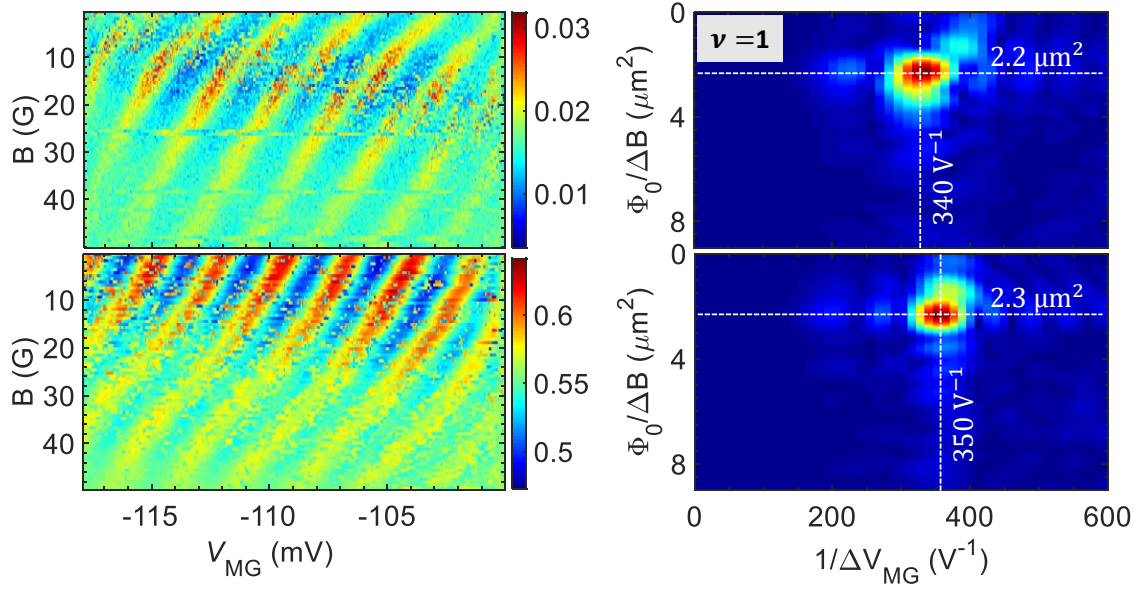

**Figure S4 | Aharonov-Bohm pajama at $\nu = 1$.** The AB interference at $\nu = 1$ ($B = 6.4$ T) for the device of area $10.25\ \mu m^2$. Top: The pajama (left) when the QPCs are strongly pinched with the transmission $t_{l,r}^2 \approx 0.14$ (i.e., below the plateau of 1/3), and the first Fourier Transformation (FFT) on the right. Bottom: Pajama and FFT when the QPCs are relatively open, $t_{l,r}^2 \approx 0.75$ (i.e., above the plateau of 1/3). The periodicities remain nearly the same with QPCs' transmission.

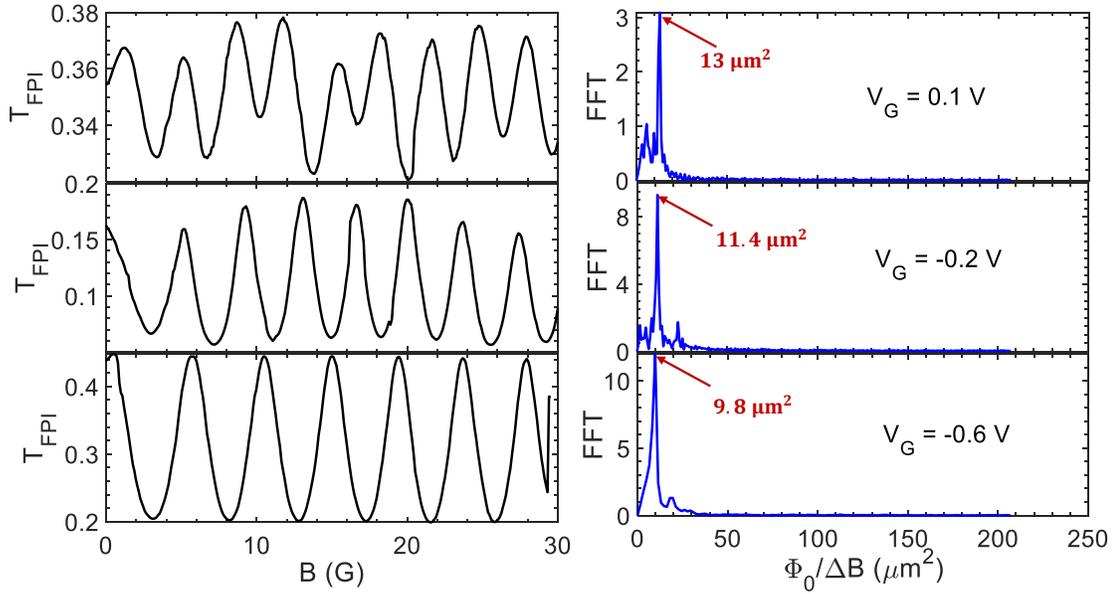

**Figure S5 | Magnetic field-dependent traces at different gate depletion voltages at $\nu = 2$.** Aharonov-Bohm oscillation of the conductance as the transmission $T_{\text{FPI}}$ as a function of the magnetic field (left) and their first Fourier transformation (right) at different gate voltages, $V_G$. The peak shows a continuous reduction in the effective area with the depletion under the gates getting stronger. The lithographic area of the device is 14.2 μm².

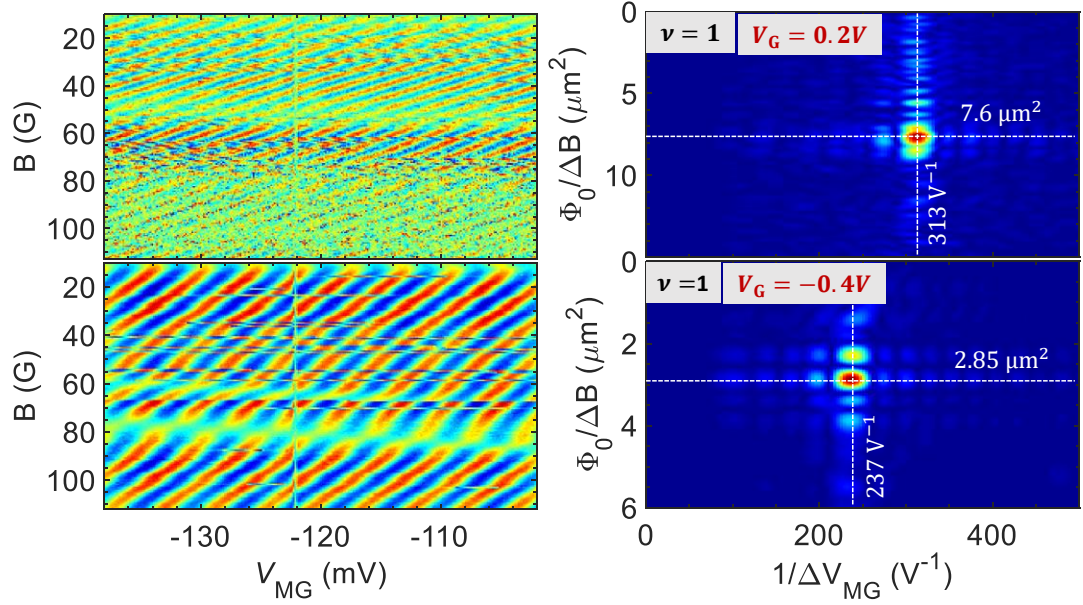

**Figure S6 | Aharonov-Bohm interference at $\nu = 1$ for a softer and a stronger gate depletion.** A typical pajama (left) and the first Fourier transformation (right) at the filling of $\nu = 1$ for two different gate confining voltages: $V_G = 0.2$ V (soft pinch) and $V_G = -0.4$ V (strong pinch). A drastic difference (~62% less) in the effective area is observed with a stronger pinch. The lithographic area of the device is $14.2\ \mu m^2$.

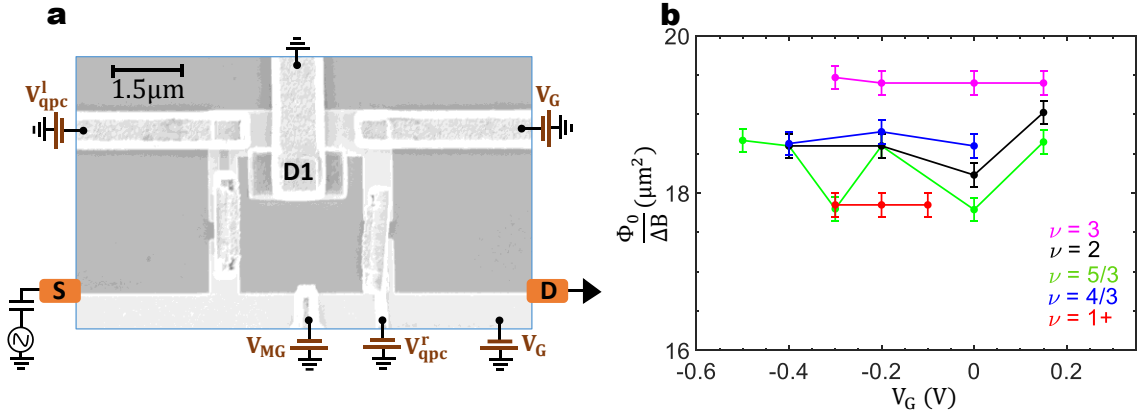

**Figure S7 | Results for a Mach-Zehnder interferometer.** (a) Scanning electron micrograph image of a Mach-Zehnder interferometer with the in-built grounded drain D1 inside the interferometer. The QPC is tuned separately as compared to the other gates. (b) The variation of $\left(\frac{\Phi_0}{\Delta B}\right)$ with gate depletion voltage $V_G$ at different filling factors shows negligible ($< 5\%$) change with both the magnetic field (filling factor) and gate depletion.

**Section-I: Detailed analysis of anomalous field periodicity due to dynamic area change in the presence of edge reconstruction.**

The flux evolution is given by $\Delta\Phi = \Delta B * A + B * \Delta A$. In a declining filling factor, the second term $B * \Delta A$ must be included at the filling $\nu \approx 5/3$ where the ubiquitous edge reconstruction is born. During magnetic field scanning from 5/3 and below, two things happen simultaneously: **i.** the conductance evolves periodically with each period of flux quantum $\Delta\Phi = \Phi_0$, and **ii.** The interfering reconstructed edge mode keeps on widening. Starting from 5/3, the consecutive $\Phi_0$ period can be written as $\Phi_0 = \Delta B_1 * A + B_1 * \Delta A_1$, $\Phi_0 = \Delta B_2 * (A - \Delta A_1) + B_2 * \Delta A_2$, $\Phi_0 = \Delta B_3 * (A - \Delta A_1 - \Delta A_2) + B_3 * \Delta A_3$, and so on. Thus, one observes a gradual increase in $\Delta B_{1,2,...}$ From the obtained field periodicities for a device (marked by blue in Fig. 3(a) of the main text), we approximately estimate the total number of flux quantum to be $\approx 1920$, when the $B$ is varied from 4.4 T ($\nu \approx 5/3$) to 6.8 T ($\nu = 1$) and considering an average $\Delta B = \frac{1}{2}(6.4 + 18.6)$ G. The variation of the calculated $\Delta A$ and the (effective) AB area $A$ with $B$ within the filling range $\nu = \frac{5}{3} - 1$ are shown in Fig. S8. Note that the area $A$ changes from 6.4 µm² to 5.33 µm² suggesting $\approx 230$ nm widening of the edge mode at 6.8 T, in contrast to the huge shrinking of the area according to the usual $A = \frac{\Phi_0}{\Delta B}$.

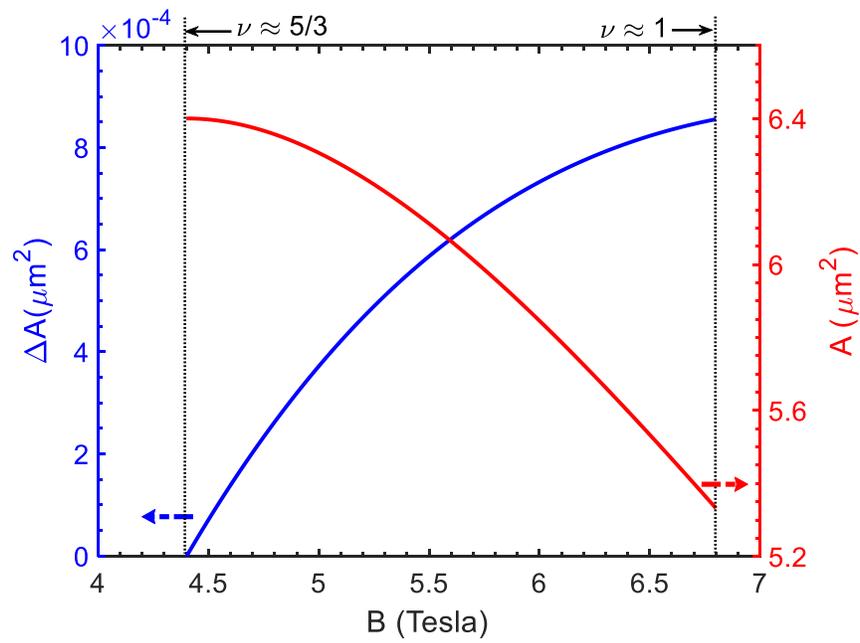

**Figure S8 | Analysis of the dynamic area between the fillings 5/3 and 1**. Estimated $\Delta A$ (blue) and the effective area $A$ (red) as a result of the continuing widening of the reconstructed edge with the magnetic field.